\documentclass[12pt,letterpaper]{article}    
\usepackage{osajnl2} 
\usepackage{amsfonts}      
\usepackage{amsmath}        
\usepackage{amssymb}        
\usepackage{amsbsy}         
\usepackage[mathscr]{eucal} 
\usepackage{graphicx}       

\begin{document}
\title{Local Normal Mode Coupling and Energy Band Splitting in Elliptically Birefringent 1D Magnetophotonic Crystals}
\author{Amir A Jalali and Miguel Levy}
\address{Michigan Technological University, Department of Physics, 1400 Townsend Dr, Houghton, MI 49931, USA}

\bibliographystyle{osajnl}

\begin{abstract}
An analysis is presented of wave-vector dispersion in elliptically birefringent stratified magneto-optic media having one-dimensional periodicity.  It is found that local normal-mode polarization-state differences between adjacent layers lead to mode coupling and impact the wave-vector dispersion and the character of the Bloch states of the system.  This coupling produces extra terms in the dispersion relation not present in uniform circularly birefringent magneto-optic stratified media.  Normal mode coupling lifts the degeneracy at frequency band cross-over points under certain conditions and induces a magnetization-dependent optical band gap.  This study examines the conditions for band gap formation in the system.  It shows that such a frequency-split can be characterized by a simple coupling parameter that depends on the relation between polarization states of local normal modes in adjacent layers.  The character of the Bloch states and conditions for maximizing the strength of the band splitting in these systems are analyzed.
\end{abstract}

\maketitle

\section{Introduction}
A number of publications have discussed the propagation of electromagnetic waves in one-dimensional birefringent stratified media.~\cite{Yeh1979,Mandatori2003} Of particular interest is the formulation pioneered by P. Yeh that employs a translation matrix approach to discuss periodic linearly birefringent structures.  The propagation of light across a single period is analyzed in conjunction with Floquet's theorem to determine the dispersion relation and Bloch waves of the system.  This approach has been used to study the properties of layered media with misaligned optic axes from one layer to the next.~\cite{Mandatori2003} Numerical solutions have been found for normal incidence of an optical beam for identical birefringent plates with alternating azimuth angles and layer thicknesses.~\cite{Yeh1979}

More recently the present authors have used this technique to study elliptically birefringent non-reciprocal media,~\cite{Levy2007} where elliptically polarized normal modes characterize the system locally.  Analytic solutions were found for light incident perpendicularly into stratified media consisting of alternating magneto-optic (MO) layers with different gyration vectors~\cite{Landau1984} and birefringence levels.~\cite{Levy2007} Adjacent layers were assumed to have their anisotropy axes aligned to each other.  This model captures some important features of one-dimensional magnetophotonic crystal waveguides, particularly the presence of magneto-optical gyrotropy and locally alternating birefringence levels.  Such systems are presently being studied for use in integrated fast optical switches and ultra-small optical isolators.~\cite{Levy2006a,Li2005,Li2005a}  Work on these systems extends prior theoretical and experimental efforts on magnetophotonic crystals in order to encompass the elliptical birefringence that often characterizes planar magnetophotonic structures.~\cite{Inoue1999,Figotin2001,Khanikaev2005,Kahl2004a,Khartsev2007}

A particularly interesting feature of periodic elliptically birefringent gyrotropic media concerns the character of the dispersion branch solutions to the Floquet theorem.  The work of Pochi Yeh cited before~\cite{Yeh1979} discusses the formation of a band gap away from the Brillouin zone boundary in linearly birefringent media with misaligned anisotropy axes.  Merzlikin et.al. have pointed out that magnetically tunable band gaps can arise in stratified media that combine circularly and linearly birefringent layers.~\cite{Merzlikin2007}  In the present work we note that magnetically tunable band gaps can also exist in elliptically birefringent stratified magneto-optic media.  A range of frequency splits is possible since local normal modes span a wide spectrum of polarization states.  The work presented here traces the origin of these band structural features to normal mode coupling arising from the simultaneous presence of gyrotropy and linear birefringence in the periodic system, as the Bloch states of the system propagate across layer boundaries.  The dispersion relation for this type of system is found to contain additional terms that describe the formation of magnetically tunable band gaps away from the Brillouin zone boundary and that determine the magnitude of the gap.  A new kind of parameter is identified herewith that characterizes the coupling of these layer-dependent normal modes.  The underlying phenomenon is the continuity of transverse electric and magnetic field components across inter-layer boundaries.  Such a coupling can yield magnetization-dependent optical band gaps inside the Brillouin zone as discussed below. The formulation presented herein can serve a tool for the design of magnetically tunable band gaps in non-reciprocal magnetophotonic structures.

After introducing the formalism to be employed in this work (Sec.~\ref{sec2}), it is shown that the transfer matrix for elliptically-birefringent magnetophotonic crystals can be parametrized in terms of an inter-modal coupling parameter (Sec.~\ref{sec3}).  The conditions for the frequency splitting of degenerate Bloch states is then discussed (Sec.~\ref{sec4}) and the dispersion relation for magneto-optic layered structures in the presence of elliptical birefringence is derived and discussed (Sec.~\ref{sec5}). Section~\ref{sec6} examines the character of Bloch states for these systems.

\section{Waves in a birefringent magnetophotonic medium}
\label{sec2}
In the optical wavelength regime, the permeability of a birefringent uniaxial magnetooptic medium is very close to the permeability of vacuum $\mu_0$, its relative permeability close to unity. The relative permittivity tensor $\tilde{\epsilon}$ of the medium for magnetization along the $z-$axis, has the form
\begin{equation}
\label{1}
    \tilde{\epsilon}=\left(
    \begin{array}{c c c}
    \epsilon_{xx} & i\epsilon_{xy} & 0\\
    -i\epsilon_{xy} & \epsilon_{yy} & 0\\
    0 & 0 & \epsilon_{zz}\\
    \end{array}
    \right),
\end{equation}
where we assume no absorption of the light in the medium. This implies that all components of the relative permittivity ($\epsilon_{i,j}, i,j=x,y,z$) are real and it is not assumed that $\epsilon_{xx}=\epsilon_{yy}$. By solving the wave equation upon normal incidence of a monochromatic plane wave (with time dependence $\exp{(i \omega t)}$) propagating parallel to the $z$ axis on a birefringent magnetooptic medium, one obtains eigenmodes
\begin{eqnarray}
 \label{2}
    \mathbf{\hat{e}}_{\pm}=\frac{1}{\sqrt{2}}
    \left(
                \begin{array}{c}
                \cos\alpha \pm \sin \alpha \\
                \pm i\cos \alpha - i \sin \alpha\\
                0\\
                \end{array}
    \right),
\end{eqnarray}
corresponding to the refractive indices
\begin{eqnarray}
n_{\pm}^2 = \bar{\epsilon} \pm  \sqrt{\Delta^2+\epsilon_{xy}^2}\;.
\end{eqnarray}
Here $\bar{\epsilon}=(\epsilon_{yy} + \epsilon_{xx})/2$, $\Delta~=~(\epsilon_{yy}-~\epsilon_{xx})/2$, and $\tan (2\alpha)=\Delta/\epsilon_{xy}$. The propagation constant in the medium is defined as \begin{eqnarray}
 \beta_{\pm} &=& \frac{\omega}{c}n_{\pm}
\end{eqnarray}
where $c$ is the speed of light in vacuum.
\section{One-dimensional birefringent magnetophotonic crystals}
\label{sec3}
\begin{figure}[t]
\centerline{\includegraphics{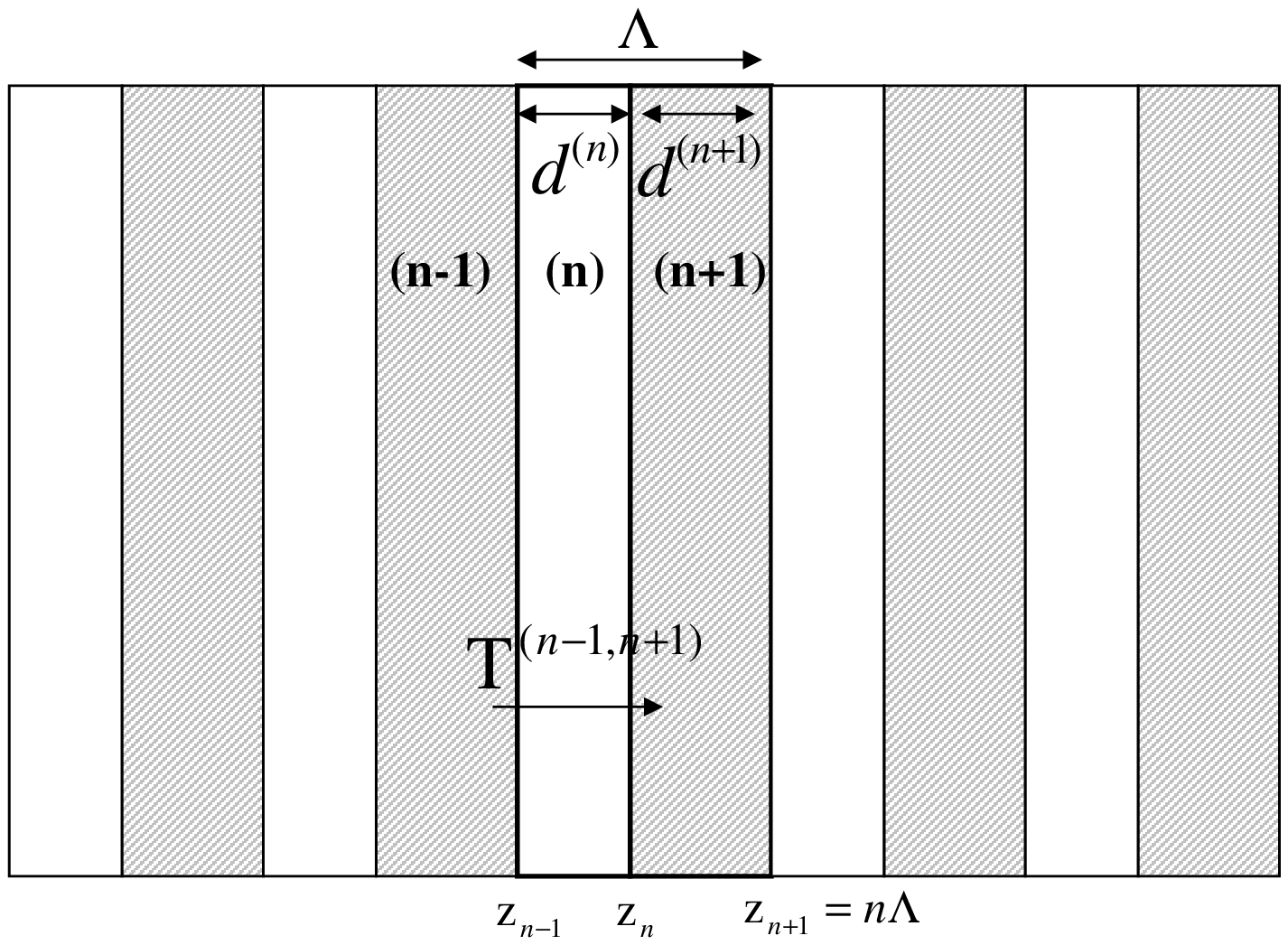}}
\caption{Schematic diagram of a one-dimensional birefringent magnetophotonic crystal with period of $\Lambda$. The magnetophotonic crystal extend indefinitely in the $x$ and $y$ directions. A plane wave is incident normally to the layered structure. jalaliF1.eps}
\label{fig1}
\end{figure}

Consider a plane wave normally incident on a periodic stack structure of alternating elliptically  birefringent magnetophotonic layers (Fig.~\ref{fig1}), where the elliptical birefringence parameters of adjacent layers may differ, but the anisotropy axis are aligned. The Bloch states for this system can be expressed in terms of local normal modes. Thus in the $n$-th layer the optical electric field can be written as
\begin{eqnarray}
\label{5}
    \mathbf{E}^{(n)}(z)&=&\left(E^{(n)}_{01} e^{i\beta^{(n)}_{+}(z-z_n)}+ E^{(n)}_{02} e^{-i\beta^{(n)}_{+}(z-z_n)}\right)\mathbf{\hat{e}_{+}}^{(n)}\\
              \nonumber &&+\left(E^{(n)}_{03} e^{i\beta^{(n)}_{-}(z-z_n)}+ E^{(n)}_{04} e^{-i\beta^{(n)}_{-}(z-z_n)}\right)\mathbf{\hat{e}_{-}}^{(n)}.
\end{eqnarray}
Here $E^{(n)}_{0i}$ $(i=1,\cdots,4)$ are the complex amplitudes of the partial waves corresponding to each normal mode. The Bloch states for this system satisfy the Floquet-Bloch theorem through the following eigenvalue equation.
\begin{eqnarray}
\label{Floquet}
  \mathbf{T}^{(n-1,n+1)}\mathbf{E} &=& \exp{(i K \Lambda)} \mathbf{E},
\end{eqnarray}
where the transfer matrix $\mathbf{T}^{(n-1,n+1)}$  relates the four eigenmode amplitudes $\mathbf{E}$ in the second layer of a unit cell (the region between $z=(n-2)\Lambda$ and $z=(n-1)\Lambda$) to the corresponding amplitudes in the second layer of the adjacent unit cell (the region between $z=(n-1)\Lambda$ and $z=n\Lambda$).  $K$ is the Bloch wave vector and $\Lambda$ is the period of the periodic structure.
With this functional basis, the transfer matrix $\mathbf{T}^{(n-1,n+1)}$ acquires the following form:
\begin{eqnarray}
\label{Tmatrix}
\nonumber
  \begin{pmatrix}
    f_{1,1}+g_{1,1}\sin^{2}\chi^{(n,n+1)} & f_{1,2}+g_{1,2} \sin^2 \chi^{(n,n+1)} & g_{1,3}\sin 2\chi^{(n,n+1)} & g_{1,4}\sin 2\chi^{(n,n+1)} \\
    f_{2,1}+g_{2,1} \sin^2 \chi^{(n,n+1)} & f_{2,2}+g_{2,2}\sin^{2}\chi^{(n,n+1)} & g_{2,3}\sin 2\chi^{(n,n+1)} & g_{2,4}\sin 2\chi^{(n,n+1)} \\
    g_{3,1}\sin 2\chi^{(n,n+1)} & g_{3,2}\sin 2\chi^{(n,n+1)} & f_{3,3}+g_{3,3}\sin^{2}\chi^{(n,n+1)} & f_{3,4}+g_{3,4} \sin^2 \chi^{(n,n+1)} \\
    g_{4,1}\sin 2\chi^{(n,n+1)} & g_{4,2}\sin 2\chi^{(n,n+1)} & f_{4,3}+g_{4,3} \sin^2 \chi^{(n,n+1)} & f_{4,4}+g_{4,4}\sin^{2}\chi^{(n,n+1)}\\
  \end{pmatrix}
  \\
\end{eqnarray}
where
\begin{eqnarray}
  f_{1,1} &=& \exp(-i \beta^{(n+1)}_{+} d^{(n+1)})\left[\cos (\beta^{(n)}_{+}d^{(n)}) - \frac{i}{2}\left(\frac{n^{(n)}_{+}}{n^{(n+1)}_{+}} + \frac{n^{(n+1)}_{+}}{n^{(n)}_{+}}\right)\sin (\beta^{(n)}_{+}d^{(n)})\right]  \\
  \nonumber g_{1,1} &=& \exp(-i \beta^{(n+1)}_{+}d^{(n+1)})\left[ \cos( \beta^{(n)}_{-}d^{(n)}) - \cos(\beta^{(n)}_{+}d^{(n)}) - \right. \\
  && \left. \frac{i}{2}\left(\frac{n^{(n)}_{-}}{n^{(n+1)}_{+}} + \frac{n^{(n+1)}_{+}}{n^{(n)}_{-}}\right)\sin(\beta^{(n)}_{-}d^{(n)})
  + \frac{i}{2}\left(\frac{n^{(n)}_{+}}{n^{(n+1)}_{+}} + \frac{n^{(n+1)}_{+}}{n^{(n)}_{+}}\right)\sin(\beta^{(n)}_{+}d^{(n)})\right]\\
  f_{1,2} &=& \frac{i}{2} \exp{(i\beta^{(n+1)}_{+}d^{(n+1)})} \left[\frac{n^{(n+1)}_{+}}{n^{(n)}_{+}} - \frac{n^{(n)}_{+}}{n^{(n+1)}_{+}}\right]\sin (\beta^{(n)}_{+}d^{(n)})\\
  \nonumber g_{1,2} &=& \frac{i}{2} \exp{(i \beta^{(n+1)}_{+}d^{(n+1)})} \times \\
  && \left[\left(\frac{n^{(n)}_{+}}{n^{(n+1)}_{+}} - \frac{n^{(n+1)}_{+}}{n^{(n)}_{+}}\right)\sin(\beta^{(n)}_{+}d^{(n)}) + \left(\frac{n^{(n+1)}_{+}}{n^{(n)}_{-}} - \frac{n^{(n)}_{-}}{n^{(n+1)}_{+}}\right)\sin(\beta^{(n)}_{-}d^{(n)})\right]\\
  \nonumber g_{1,3} &=& \exp{(-i \beta^{(n+1)}_{-}d^{(n+1)})} \left[\left(\frac{n^{(n+1)}_{-}+n^{(n+1)}_{+}}{4 n^{(n+1)}_{+}}\right)\left(\cos(\beta^{(n)}_{+}d^{(n)}) - \cos(\beta^{(n)}_{-}d^{(n)})\right) + \right.\\
  \nonumber && \left.\frac{i}{4}\left(\frac{n^{(n+1)}_{-}}{n^{(n)}_{-}} + \frac{n^{(n)}_{-}}{n^{(n+1)}_{+}}\right)\sin(\beta^{(n)}_{-}d^{(n)}) - \frac{i}{4}\left(\frac{n^{(n+1)}_{-}}{n^{(n)}_{+}} + \frac{n^{(n)}_{+}}{n^{(n+1)}_{+}}\right)\sin(\beta^{(n)}_{+}d^{(n)})\right]\\
  \end{eqnarray}
  \begin{eqnarray}
  \nonumber g_{1,4} &=& \exp{(i \beta^{(n+1)}_{-}d^{(n+1)})} \left[\left(\frac{n^{(n+1)}_{+}-n^{(n+1)}_{-}}{4 n^{(n+1)}_{+}}\right)\left(\cos(\beta^{(n)}_{+}d^{(n)}) - \cos(\beta^{(n)}_{-}d^{(n)})\right) - \right.\\
  \nonumber && \left.\frac{i}{4}\left(\frac{n^{(n+1)}_{-}}{n^{(n)}_{-}} - \frac{n^{(n)}_{-}}{n^{(n+1)}_{+}}\right)\sin(\beta^{(n)}_{-}d^{(n)}) + \frac{i}{4}\left(\frac{n^{(n+1)}_{-}}{n^{(n)}_{+}} - \frac{n^{(n)}_{+}}{n^{(n+1)}_{+}}\right)\sin(\beta^{(n)}_{+}d^{(n)})\right].\\
  \end{eqnarray}
Here $d^{(n)}$ is the thickness of layer $n$ in the stack and $\chi^{(n,n+1)}=\alpha^{(n)}-\alpha^{(n+1)}$. The other elements of the $\mathbf{T}^{(n-1,n+1)}$ matrix can be obtained by the following symmetry relations
\begin{eqnarray}
\nonumber f_{2,1} &=& f_{1,2}^{\ast} \hspace{0.5 cm};\hspace{0.5 cm} g_{2,1}=g_{1,2}^{\ast}\\
\nonumber f_{2,2} &=& f_{1,1}^{\ast} \hspace{0.5 cm};\hspace{0.5 cm} g_{2,2}=g_{1,1}^{\ast}\\
\nonumber g_{2,3} &=& g_{1,4}^{\ast}\\
\nonumber g_{2,4} &=& g_{1,3}^{\ast}\\
\nonumber g_{3,1} &=& g_{1,3} \hspace{3.9 cm} (n^{(n+1)}_{+} \leftrightarrow n^{(n+1)}_{-})\\
\nonumber g_{3,2} &=& g_{1,3} \hspace{3.9 cm} (n^{(n+1)}_{+} \leftrightarrow n^{(n+1)}_{-})\\
\nonumber f_{3,3} &=& f_{1,1} \hspace{0.5 cm};\hspace{0.5 cm} g_{3,3} = g_{1,1} \hspace{1 cm} (n^{(n+1)}_{+} \leftrightarrow n^{(n+1)}_{-} \mbox{~and~} n^{(n)}_{+} \leftrightarrow n^{(n)}_{-})\\
\nonumber f_{3,4} &=& f_{1,2} \hspace{0.5 cm};\hspace{0.5 cm} g_{3,4} = g_{1,2} \hspace{1 cm} (n^{(n+1)}_{+} \leftrightarrow n^{(n+1)}_{-} \mbox{~and~} n^{(n)}_{+} \leftrightarrow n^{(n)}_{-})\\
\nonumber g_{4,1} &=& g_{3,2}^{\ast} \\
\nonumber g_{4,2} &=& g_{3,1}^{\ast} \\
\nonumber f_{4,3} &=& f_{3,4}^{\ast}\hspace{0.5 cm};\hspace{0.5 cm} g_{4,3}=g_{3,4}^{\ast}\\
          f_{4,4} &=& f_{3,3}^{\ast}\hspace{0.5 cm};\hspace{0.5 cm} g_{4,4}=g_{3,3}^{\ast}
\end{eqnarray}
The $\leftrightarrow$ sign means the exchange of parameters $n^{(n,n+1)}_{+}$ with $n^{(n,n+1)}_{-}$.

\section{Normal mode coupling and the splitting of degenerate Bloch states}
\label{sec4}
Pochi Yeh has discussed the propagation of electromagnetic waves in alternating linearly birefringent layers where the normal modes differ in adjacent layers due to anisotropy axes misalignment.~\cite{Yeh1979}  He points out that in this case a new type of constructive interference in the scattered waves arises from the coupling between slow and fast waves.  Forbidden frequency zones or band gaps can now appear away from the Brillouin zone boundaries.  He calls this an exchange Bragg condition because forward propagating fast (slow) Bloch states couple to backwards propagating slow (fast) Bloch states.  A similar type of constructive interference exists in elliptically birefringent magneto-optic layered systems with different local modes in adjacent layers. It is shown here that such local normal mode variations lead to the opening up of a band gap in the Brillouin zone.  This effect can be traced to the presence of normal mode coupling between adjacent layers.

Notice that the unit cell transformation matrix $\mathbf{T}^{(n-1,n+1)}$, Eq.~(\ref{Tmatrix}), depends on the relative elliptical birefringence parameter  $\chi^{(n,n+1)}$ and the individual normal mode propagation constants for each layer.  When the parameter $\chi^{(n,n+1)}$ equals zero the normal modes in adjacent layers are the same and they remain uncoupled.  This can be seen explicitly from the form of the $\mathbf{T}^{(n-1,n+1)}$ matrix formulated in terms of normal modes.  It is clear from the expression that if $\chi^{(n,n+1)}=0$ the off-block-diagonal components of the $\mathbf{T}^{(n-1,n+1)}$ matrix are zero and there is no admixture of the local normal modes.  This situation is similar to the case of a periodic layered medium consisting of isotropic layers, where transverse electric (TE) and transverse magnetic (TM) waves, or right- and left-circularly polarized waves remain uncoupled.~\cite{Khanikaev2005}

When $\chi^{(n,n+1)}$ differs from zero the Bloch states require an admixture of local normal modes, since the off-block-diagonal components of $\mathbf{T}^{(n-1,n+1)}$ differ from zero.  Moreover, the strength of the coupling, parametrized as the weight of the off-block diagonal terms, can be seen to increase as $\sin 2\chi^{(n,n+1)}$.  We thus see that $\chi^{(n,n+1)}$ parametrizes the degree of admixture of the normal modes.  In the next section we shall discuss the form taken by this coupling in the wave-vector dispersion of the system.

As $\chi^{(n,n+1)}$ changes away from zero the polarization state of the normal modes changes according to Eq.~(\ref{2}).  Local normal modes in adjacent layers acquire different polarization states.  These normal modes in adjacent layers are now coupled by the continuity of tangential components of the magnetic and electric fields across the boundary.  Changes in normal mode polarization thus affect the solution to the Floquet-Bloch theorem through mode coupling across the boundary.

This effect can be seen in Fig.~(\ref{fig2}) below.  The new band gap that develops away from the Brillouin zone edge will be denoted as Yeh band gap since it was first noted by P.~Yeh for the case of linearly birefringent stacks.  Although the case we are considering differs from Yeh's treatment due to the elliptical birefringence and non-reciprocity of the magneto-optic systems under consideration, it is still the presence of birefringence that is the key to the formation of this new type of gap.

\section{Dispersion relation as a function of inter-modal coupling parameter}
\label{sec5}
In this section we explicitly find the dependence of the dispersion relation on the inter-modal coupling parameter $\chi^{(n,n+1)}$.  The derivation and final form of the dispersion relation highlight the $\chi^{(n,n+1)}$ dependence. We will consider the polarization state dependence on  $\chi^{(n,n+1)}$ in the next section.

If one maintains the same index contrast in the photonic crystal and the same normal mode propagation constants in each layer but allows the polarization state of the normal modes to vary, the only parameter that changes in the $\mathbf{T}^{(n-1,n+1)}$ matrix is $\chi^{(n,n+1)}$, and the band splitting that emerges is a result of normal mode coupling.

The characteristic equation of the $\mathbf{T}^{(n-1,n+1)}$ matrix in Eq.~(\ref{Tmatrix}) is a polynomial function of $\lambda$ of order four.  $\lambda$ denotes the eigenvalue $\exp{(i K \Lambda)}$ of the Floquet-Bloch expression Eq.~(\ref{Floquet}).  It does not denote wavelength as is otherwise customary. For MO materials the $g_{ij}$'s are much smaller than $f_{i,j}$'s in the $\mathbf{T}^{(n-1,n+1)}$ matrix. The characteristic equation can be written as:
\begin{eqnarray}
  \nonumber g(\lambda) &=& f(\lambda) + h(\lambda) \sin^2 \chi^{(n,n+1)} +j(\lambda) \sin^2 2\chi^{(n,n+1)}  \\
  && + k(\lambda) \sin^2 3\chi^{(n,n+1)} + l(\lambda) \sin^2 4\chi^{(n,n+1)} =0,
\end{eqnarray}
where $f(\lambda)$ is the characteristic equation of $\mathbf{T}^{(n-1,n+1)}$ matrix when $\chi^{(n,n+1)}=0$, $h$, $j$, $k$, and $l$ are functions of $n^{(n,n+1)}_{\pm}$, $\beta^{(n,n+1)}_{\pm}$, and $d^{(n,n+1)}$. We note that for an MO periodic structure,  $f\gg h \gg j\gg k \gg l$. The zeros of $f(\lambda)$ are denoted by $\lambda'_0$ (which corresponds to the case where the off-block diagonal elements of $\mathbf{T}^{(n-1,n+1)}$ equal zero) and of $g(\lambda)$ by $\lambda_0$. Upon first order expansion of $g(\lambda)$ around $\lambda'_0$ we obtain
\begin{eqnarray}
  \nonumber g(\lambda)&=& g(\lambda'_0) + g'(\lambda'_0) (\lambda-\lambda'_0)\\
  &\approx& h(\lambda'_0) \sin^2\chi^{(n,n+1)} + (f'(\lambda'_0) + h'(\lambda'_0) \sin^2\chi^{(n,n+1)})(\lambda-\lambda'_0).
  \end{eqnarray}
  From this expression one can obtain $\lambda_0$ in terms of $\lambda'_0$
  \begin{eqnarray}
  \label{expansion}
  \nonumber (\lambda_0-\lambda'_0)&=& - \frac{h(\lambda'_0)\sin^2 \chi^{(n,n+1)}}{f'(\lambda'_0)+h'(\lambda'_0)\sin^2\chi^{(n,n+1)}}\\
  \lambda_0& \approx & \lambda'_0- \frac{h(\lambda'_0)}{f'(\lambda'_0)}\sin^2 \chi^{(n,n+1)},
\end{eqnarray}
where a prime on the functions means the first derivative with respect to their argument. In Eq.~(\ref{expansion}) use has made of the fact that $g(\lambda_0)=0$ and $f(\lambda'_0)=0$. Terms proportional to $j$, $k$, and $l$ and their derivatives have been neglected as these are all small polynomial expressions in $\lambda$. From this simple analysis we can see that the difference in eigenvalues is directly proportional to $\sin^2\chi^{(n,n+1)}$.

Let us now consider the dispersion relation Eq.~(35) in Ref.~\onlinecite{Levy2007} for the case where $\chi^{(n,n+1)}=0$, given by
\begin{eqnarray}
 \nonumber \cos K_{\pm} \Lambda &=&  \left(\cos (\beta^{(n+1)}_{\pm}d^{(n+1)}) \cos (\beta^{(n)}_{\pm}d^{(n)})-\frac{1}{2} N_{\pm} \sin (\beta^{(n+1)}_{\pm}d^{(n+1)}) \sin (\beta^{(n)}_{\pm}d^{(n)})\right).\\
\end{eqnarray}
In this case band gaps appear only at the boundary of the Brillouin zone, displaying complex solutions for $K \Lambda$.  On the other hand, when  $\chi^{(n,n+1)}$ differs from zero, the dispersion relation acquires an additional term in the form of $\Re{[h(\lambda'_0)/f'(\lambda'_0)]}\sin^2 \chi^{(n,n+1)}$, as follows:
\begin{eqnarray}
\label{dispersion}
 \nonumber \cos K_{\pm} \Lambda &=&  \left(\cos (\beta^{(n+1)}_{\pm}d^{(n+1)}) \cos (\beta^{(n)}_{\pm}d^{(n)})-\frac{1}{2} N_{\pm} \sin (\beta^{(n+1)}_{\pm}d^{(n+1)}) \sin (\beta^{(n)}_{\pm}d^{(n)})\right)\\
 &&-\Re{\left(\frac{h(\lambda'_{0_{\pm}})}{f'(\lambda'_{0_{\pm}})}\right)}\sin^2 \chi^{(n,n+1)}.
\end{eqnarray}
Here $\Re$ denotes the real part of its argument and $N_{\pm}=n^{(n)}_{\pm}/n^{(n+1)}_{\pm}+n^{(n+1)}_{\pm}/n^{(n)}_{\pm}$. For a general MO material $N_{\pm}\approx 2$. This extra term in the dispersion relation is responsible for band gap formation away from the Brillouin zone edges. A complex solution for $K_{+} \Lambda$ (the same treatment can be applied for $K_{-}\Lambda$), and hence the existence of a band gap, occurs under the following conditions
\begin{eqnarray}
\label{Kcondition}
  \nonumber \cos 2\bar{\beta}\Lambda & < & \frac{ 4+4u(\lambda'_{0_{+}})\sin^2 \chi^{(n,n+1)}}{2+N_+}\\
  \cos 2\bar{\beta}\Lambda & > & \frac{-4+4u(\lambda'_{0_{+}})\sin^2 \chi^{(n,n+1)}}{2+N_+},
\end{eqnarray}
where $\bar{\beta} \Lambda=\omega/(2 c)(n^{(n)}_{+} d^{(n)}+n^{(n+1)}_{+} d^{(n+1)})$ and $u(\lambda'_{0_{+}})=\Re (h(\lambda'_{0_{+}})/f'(\lambda'_{0_{+}}))$.

If one maintains the same refractive index contrast and normal mode propagation constants for each layer of the periodic structure and allows the normal mode polarizations to change, the Yeh band gap width is a function of $\chi^{(n,n+1)}$ only.  This band width increases monotonically with  $0 < \chi^{(n,n+1)} < \pi/2$, as the range of $\bar{\beta} \Lambda$ expands according to Eq.~(\ref{Kcondition}).  From Eq.~(\ref{dispersion}) $K$  acquires a complex solution only for negative $u(\lambda'_{0_{+}})$ for a typical MO periodic structure. The upper bound of $\bar{\beta}\Lambda$ will thus have larger values and the lower bound lower values as  $\chi^{(n,n+1)}$ increases. Notice that these bounds occur under equality in Eq~(\ref{Kcondition}). This results in a wider Yeh band gap.

As an example let us consider a model system composed of bismuth iron garnet (BiIG) with typical values for the dielectric tensor in the near infrared region
\begin{equation}
    \tilde{\epsilon}^{(n)}=\left(
    \begin{array}{c c c}
    6.5411 & i0.018 & 0\\
    -i0.018 & 6.611 & 0\\
    0 & 0 & \epsilon_{zz}\\
    \end{array}
    \right)\mbox{ and }
    \tilde{\epsilon}^{(n+1)}=\left(
    \begin{array}{c c c}
    5.9859 & i0.018 & 0\\
    -i0.018 & 6.1699 & 0\\
    0 & 0 & \epsilon_{zz}\\
    \end{array}
    \right).
        \end{equation}
 This structure simulates the effective index variation of a magnetophotonic crystal on ridge waveguide.~\cite{Huang2006} In Fig.~\ref{fig2}, we show the band structure for the case where $\chi^{(n,n+1)}=0$ in dashed lines, corresponding to the solutions for $f(\lambda)$. There is no band splitting in the band structure at the crossover point. On the other hand, when $\chi^{(n,n+1)}\neq 0$, a band gap opens up, as shown by the solid lines.

In Fig.~\ref{fig3} we show the variation of the Yeh gap bandwidth for the same structure upon variation of the off-diagonal components in the dielectric tensors, corresponding to the tuning of magnetization by an external magnetic field applied to the photonic crystal structure. We also show in the same figure the variation of the Yeh gap bandwidth upon the variation of birefringence in the waveguide periodic structure. We maintain the same refractive index contrast in the periodic structure while the birefringence of the layers varies. In both cases the Yeh gap bandwidth increases with $\chi^{(n,n+1)}$.

\section{Bloch states in periodic birefringent media}
\label{sec6}
Let us define the $n$-th unit cell as the combination of layers of $n$ and ($n+1$). The translation matrix from the second layer of the ($n-1$)-th unit cell to the second layer of $n$-th unit cell is given by the $\mathbf{T}^{(n-1,n+1)}$ matrix in Eq.~(\ref{Tmatrix}). Upon solving the eigenvalue equation~(\ref{Floquet}), the eigenvectors are given by
\begin{eqnarray}
 \mathbf{E}_K=c_K
    \begin{pmatrix}
     A_n \\
     B_n\\
     C_n \\
     D_n\\
     \end{pmatrix}_K
 \end{eqnarray}
 where $c_K$ is an arbitrary constant. $A_n$, $B_n$, $C_n$, and $D_n$ are functions of $n^{(n,n+1)}_{\pm}$ , $\beta^{(n,n+1)}_{\pm}$, and $\chi^{(n,n+1)}$. According to Eq.~(\ref{5}), the Bloch wave solution in the second layer of the $n$-th unit cell is then given (up to a constant factor) by
\begin{eqnarray}
\label{polarization}
    \mathbf{E}^{(n+1)}_K(z)&=&\left(A_n e^{i\beta^{(n+1)}_{+}(z-n \Lambda)}+ B_n e^{-i\beta^{(n+1)}_{+}(z-n \Lambda)}\right)e^{i K \Lambda} \mathbf{\hat{e}_{+}}^{(n+1)}\\
    \nonumber && +\left(C_n e^{i\beta^{(n+1)}_{-}(z-n \Lambda)}+ D_n e^{-i\beta^{(n+1)}_{-}(z-n \Lambda)}\right)e^{i K \Lambda}\mathbf{\hat{e}_{-}}^{(n+1)}.
\end{eqnarray}
Note that this expression depends on the relative birefringence parameter $\chi^{(n,n+1)}$ through the components of the matrix $\mathbf{T}^{(n-1,n+1)}$. Figure~\ref{fig4} shows the polarization states of the Bloch eigenmode at the interface between layer ($n-1$) and layer $n$ and at the interface between layer $n$ and layer ($n+1$) for artificially large values of the off-diagonal component of the dielectric tensors, in order to highlight the polarization variation of the Bloch state. In this example, we have taken the material dielectric tensors as
\begin{equation}
    \tilde{\epsilon}^{(n)}=\left(
    \begin{array}{c c c}
    4 & i0.1 & 0\\
    -i0.1 & 2.5 & 0\\
    0 & 0 & \epsilon_{zz}\\
    \end{array}
    \right)\mbox{ and }
    \tilde{\epsilon}^{(n+1)}=\left(
    \begin{array}{c c c}
    6 & i2 & 0\\
    -i2 & 5 & 0\\
    0 & 0 & \epsilon_{zz}\\
    \end{array}
    \right).
        \end{equation}
In general, the polarization state of the Bloch mode will also evolve as the wave transverses the unit cell, although for typical values of the dielectric tensors these changes will be small. Thus the character of the Bloch mode is affected by the coupling.

Whereas Bloch states for non-elliptically birefringent gyrotropic one-dimensional stacks are still circularly polarized, elliptically birefringent stacks can have Bloch modes whose polarization states differ from that of the normal modes in each layer, and that depend on $\chi^{(n,n+1)}$ according to Eq.~(\ref{polarization}).

\section{Conclusions}
Mode coupling as a result of periodic variations in the polarization state of local normal modes in elliptically birefringent non-reciprocal periodic structures is reported and discussed.  Interlayer normal mode coupling in such media affects the polarization state of the Bloch waves and the wave-vector frequency dispersion.  This interlayer coupling is absent in isotropic and uniform circularly-birefringent periodic media.  As a consequence of local normal mode coupling, extra terms appear in the dispersion relation characterizing the formation of a frequency band gap inside the Brillouin zone away from the zone boundary.  The band width of this gap is found to be parametrized by a single characteristic coupling constant, and is shown to increase monotonically with this coupling parameter.  An expression for the latter is presented and shown to depend on the difference in diagonal components of the dielectric tensors and the gyrotropies of adjacent layers in the periodic structure.  Thus a ready made tool for designing wavelength dependent band gaps in non-reciprocal periodic magnetophotonic structures and calibrating their band widths is presented.  Bloch mode polarization states are found to differ from those of the local normal modes and to evolve into different elliptical states as the wave propagates down the crystal.  These Bloch mode polarization states are found to depend on the strength of the coupling between local normal modes in different layers.

\section*{Acknowledgments}
This material is based upon work supported by the National Science Foundation under Grants No. ECCS-0520814 and DMR-0709669.


\newpage

 \begin{figure}[t]
\centerline{\includegraphics{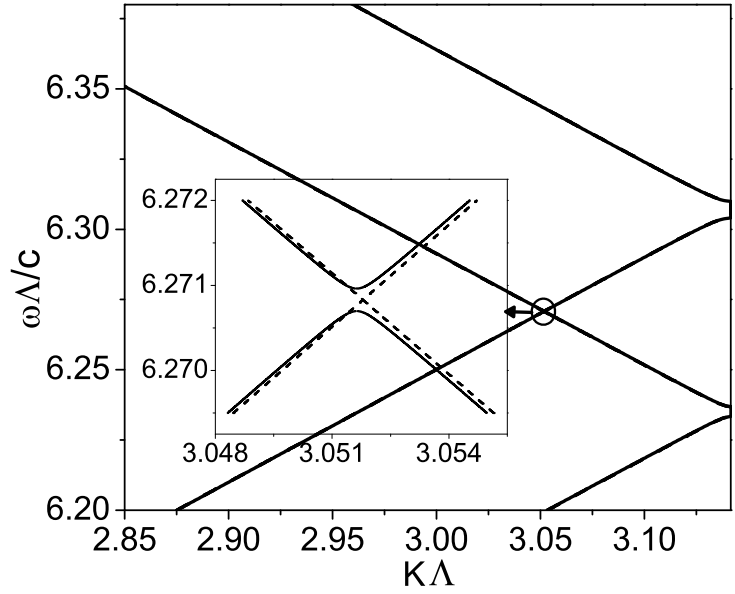}}
\caption{The energy band diagram for BiIG periodic structure (fifth branch) with $d^{(n)}=0.4$ and $d^{(n+1)}=0.6$. The dashed line corresponds to the case where $\chi^{(n,n+1)}=0$ in the transfer matrix. The solid line corresponds to a realistic case with $\chi=0.14$. jalaliF2.eps}
\label{fig2}
\end{figure}

 \begin{figure}[t]
\centerline{\includegraphics{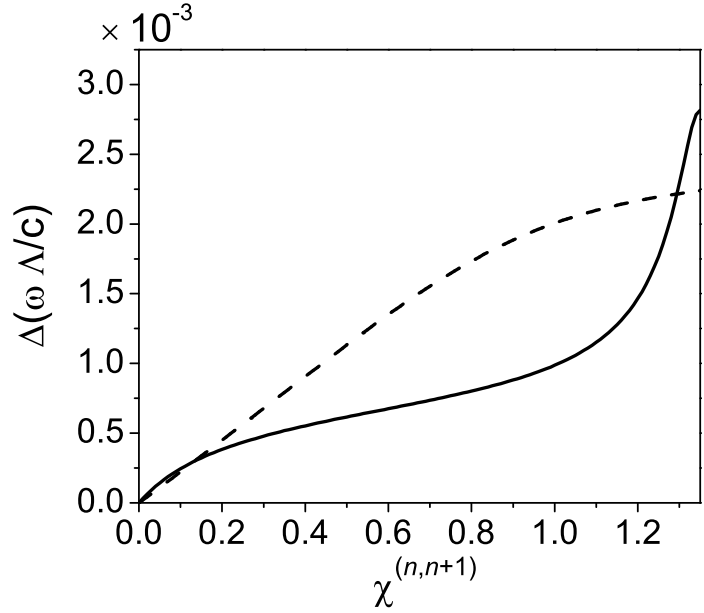}}
\caption{The width of the Yeh band gap versus $\chi^{(n,n+1)}$. The Yeh band gap was calculated for the fifth branch of the band structure of the periodic structure of BiIG with $d^{(n)}=0.4$ and $d^{(n+1)}=0.6$. In one case the average in the refractive indices is kept constant while $\chi^{(n,n+1)}$ is allowed to change (solid line). In the other, the diagonal elements of the dielectric tensors of adjacent layers are kept constant while the off-diagonal elements are allowed to change simultaneously (dashed line). jalaliF3.eps}
\label{fig3}
\end{figure}

\begin{figure}[t]
\centerline{\includegraphics{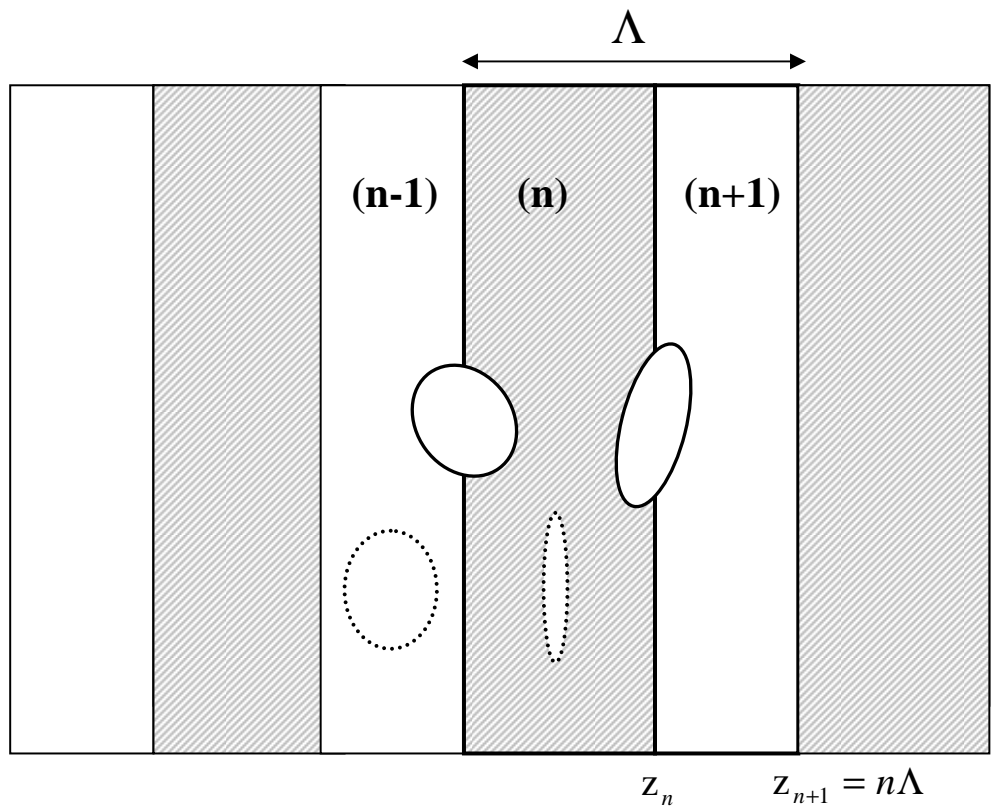}}
\caption{Polarization of Bloch wave traveling through a magnetooptic birefringent periodic structure. The Bloch wave polarization is depicted on the boundary of each layer in a unit cell (solid ellipses) just before the Yeh band gap in the band structure of the medium. Dashed ellipses show the local eigenmodes polarizations $\hat{\mathbf{e}}_+$ for each layer. jalaliF4.eps}
\label{fig4}
\end{figure}

\end{document}